\documentclass[twocolumn,showpacs,preprintnumbers,amsmath,amssymb,unsortedaddress]
{revtex4-1}

\usepackage{graphicx}
\usepackage{dcolumn}
\usepackage{bm}

\begin{document}

\preprint{AIP/123-QED}

\title{Alfv\'en seismic vibrations
of crustal solid-state plasma in quaking paramagnetic neutron star}

\author{S.  Bastrukov}\altaffiliation[Also at ]{Joint Institute for Nuclear Research, Dubna, Russia}

\affiliation{State Key Laboratory of Nuclear Physics, Peking University, Beijing, China}

\author{I. Molodtsova}
\affiliation{Joint Institute for Nuclear Research, Dubna, Russia}

\author{J. Takata}
\affiliation{
Hong Kong University, Hong Kong, China}

\author{H.-K. Chang}
\affiliation{National Tsing Hua University, Hsinchu, Taiwan}

\author{R.-X. Xu}
\affiliation{State Key Laboratory of Nuclear Physics, Peking University, Beijing, China}

\date{\today}

\begin{abstract}
 Magneto-solid-mechanical model of two-component, core-crust,
 paramagnetic neutron star responding to quake-induced
 perturbation by differentially rotational, torsional, oscillations of
 crustal electron-nuclear solid-state plasma about axis of magnetic field frozen
 in the immobile paramagnetic core is developed.
 Particular attention is given to the node-free
 torsional crust-against-core vibrations under combined action of Lorentz magnetic
 and Hooke's elastic forces; the damping is attributed to Newtonian force of
 shear viscose stresses in crustal solid-state plasma. The  spectral formulae for the frequency and lifetime of this toroidal mode are derived in analytic form and discussed in the context of quasi-periodic oscillations of the X-ray outburst
 flux from quaking magnetars. The application of obtained theoretical spectra
 to modal analysis of available data on frequencies of oscillating outburst emission suggests  that  detected
 variability is the manifestation of crustal Alfv\'en's seismic vibrations restored by Lorentz force of magnetic field stresses.
\end{abstract}


\maketitle

\section{Introduction}
  The investigations of neutron star seismic vibrations
  offer unique opportunity of studying their internal structure,
  solid-mechanical and electrodynamical properties of superdense
  degenerate matter. The most conspicuous feature of these non-convective solid
  stars is the capability of accommodating magnetic fields of extremely
  high intensity\cite{Ch-92} that serve as a chief promoter of their
  observable electromagnetic activity. The absence of nuclear energy
  sources in these final stage (FS) stars  suggests that their magnetic fields are
  definitely not generated by persistent current-carrying flows in self-exciting dynamo
  processes, as is the case of liquid main-sequence (MS) stars.  It seems quite likely,
  therefore, that stability to spontaneous decay\cite{Bh-02} of fossil magnetic fields 
  of  
  isolated neutron stars\cite{FW-04}
  is maintained by permanent magnetization of neutron-dominated (poorly
  conducting) degenerate Fermi-matter. Such an understanding has been laid at the 
  base of paramagnetic neutron star model\cite{B-01,B-02a,B-02b,B-03}. In this 
  model the degenerate Fermi-matter of non-relativistic neutrons
   (whose degeneracy pressure withstands the pressure of self-gravity) is regarded
   as being in the permanently magnetized state of field-induced Pauli's paramagnetic
   saturation which is characterized by alignment of spin magnetic moments of
   neutrons along the axis of frozen-in magnetic field.
   The most striking dynamical manifestation of spin paramagnetic polarization
   of non-conducing neutron matter is that such a matter can
    transmit perturbations by transverse magneto-mechanical waves;
    in such a wave the vector-fields of magnetization and  material
    displacements undergo coupled differentially rotational vibrations traveling along
    the axis of magnetic field.
    In a spherical mass of paramagnetic neutron star this unique feature of
    field-induced spin magnetic polarization of neutron matter is manifested in
    that such a star can undergo solely torsional vibrations about axis of its dipole
    magnetic  moment.  Based on this finding, it was argued in above works
    that the model of paramagnetic neutron star executing
    torsional axisymmetric vibrations, weakly damped by nuclear matter viscosity,
    is able to explain long periodic ($[5<P<12]$ s -- non-typical to
    young neutron stars) pulsed character of magnetar radiation
    (both, Soft Gamma Repeaters and Anomalous X-ray Pulsars) in seismically
    quiescent regime of their emission,  as being produced by torsional vibrations,
    rather than rotation as is the case of radio pulsars.

    Recent years have seen a resurgence of interest in torsional vibrations of
     magnetars,  prompted by observations\cite{I-05,WS-06,T-06} of quasi-periodic
     oscillations (QPOs) during the outburst flare from SGR 0525-66, SGR
    1806-20, and SGR  1900+14. The statistics of X-ray burst 
    of SGRs exhibits typical for earthquakes features\cite{Eps-96}.
    It is believed, therefore, that detected QPOs are of seismic origin.
    Particular attention in this development of magnetar asteroseismology has been
    paid to the following set of data on QPO frequencies\cite{M-08}
  \begin{eqnarray}
  \label{e1.1}
  && \mbox{\rm SGR}\, 1806-20: 18, 26, 29, 92, 150, 625, 1840;\\
  \label{e1.2}
  && \mbox{\rm SGR}\, 1900+14:  28, 54, 84, 155\, \mbox{\rm [Hz]}.
 \end{eqnarray}
  The corresponding periods are substantially shorter than the
  above mentioned periods of seismically quiescent pulsed emission.
  In works\cite{B-07a,B-07b,B-08,B-09a,B-09b} motivated by this discovery,
  several models of post-quake vibrational relaxation of above magnetars have been
  investigated. Particular
  attention has been given to the regime of node-free or nodeless shear axisymmetric
  vibrations. This regime is interesting in its own
  right because such vibrations have been and still are poorly investigated
  in theoretical asteroseismology of both solid FS stars and such solid celestial
  objects as Earth-like planets\cite{LW-95,AR-02}.
  In particular, in works\cite{B-07a,B-07b,B-08}, a case of the elastic-force-driven
  nodeless shear oscillations, both torsional
  --  $_0t_\ell$ and spheroidal -- $_0s_\ell$, entrapped in the crust of finite depth
  $\Delta R$ has been studied in some details with remarkable inference that dipole
  overtones of spheroidal and torsion vibrations of crust
  against immobile core exhibit features generic to Goldstone soft modes. On the
  other hand, one can probably cast doubt on arguments of the model
  presuming  the dominant role of solid-mechanical Hooke's force of
  elastic stresses because such interpretation rests on poorly justifiable
  assumption about dynamically passive role of an ultra strong magnetic field,
  that is, that the field frozen in the star remains
  unaltered in the process of vibrations.  Bearing this in mind and assuming
  that the presence of charged particles in neutron-dominated stellar matter
  imparts to it the properties of electric conductor, in works\cite{B-09a,B-09b}, the 
  post-quake relaxation of above magnetars has been studied in the model of perfectly
  conducting solid star executing global torsional vibrations
  restored by joint action of Lorentz force of magnetic field
  stresses and Hooke's force of solid-mechanical elastic stresses.
  It was found that such a model provides fairly reasonable account of general trends
  in QPO frequencies for all data from SGR 1900+14 and for SGR
  1900+14 from the range $30\leq \nu \leq 200$ Hz, but faces serious difficulties in
  interpreting low-frequency vibrations with $\nu=18$ and $\nu=26$ Hz
  in data from SGR 1806-20. Also, the model of global torsional vibrations
  leaves some uncertainties regarding the nature of
  vibrations with $\nu=625$ and  $\nu=1840$ Hz. This last issue has been
  scrutinized in recent work\cite{B-09b} from the standpoint of a solid
  star model with non-homogeneous poloidal magnetic field of well-known
  Ferraro's form. And it was found that these high-frequency QPOs can be
  properly explained as being produced by very high overtones of node-free
  torsional Alfv\'en oscillations.

 \begin{figure}[ht]
\centering{\includegraphics[width=8.0cm]{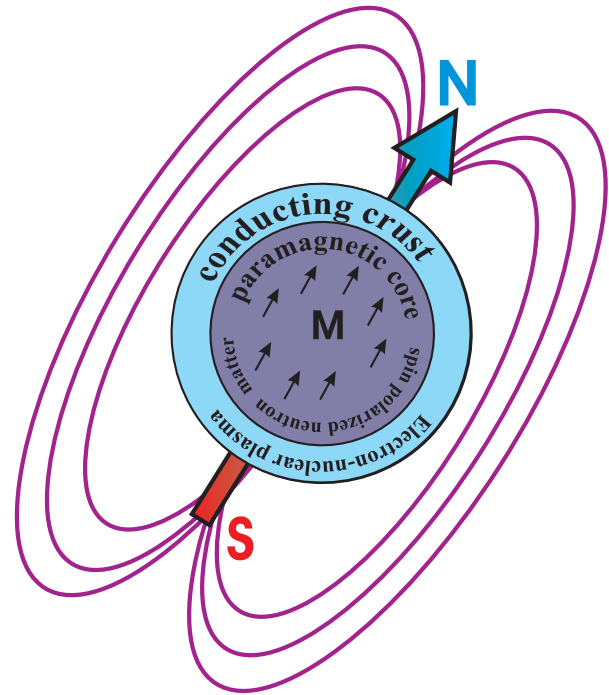}}
\caption{(Color online)
 The internal constitution of two-component, core-crust, model of paramagnetic
 neutron star. The massive core is considered as a poorly conducting permanent
 magnet composed of degenerate Fermi-gas of non-relativistic neutrons in the state
 of Pauli's paramagnetic saturation caused by field-induced alignment of spin
 magnetic moment of neutrons along the axis of uniform internal and dipolar
 external magnetic field frozen in the star on the stage of gravitational collapse of its
 MS progenitor. A highly conducting metal-like material of the neutron star crust,
 composed of nuclei embedded in the super dense degenerate Fermi-gas  of
 relativistic electrons, is regarded as  electron-nuclear solid-state magneto-active
 plasma capable of sustaining Alfv\'en oscillations.}
\end{figure}

  As a logical extension of above line of investigation, in this paper
  we consider in some details a case of  node-free torsional vibrations locked in the 
  crust with focus on toroidal Alfv\'en mode. 
  In so doing we work from the two-component model of
  paramagnetic neutron star, pictured in Fig.1, whose crust and core materials
  are regarded as endowed with substantially different electrodynamic properties.
  The immobile massive core, primarily consisting of degenerate neutron matter in 
  the above described permanently magnetized state of Pauli's paramagnetic 
  saturation, is
  regarded as a main source of  magnetic field of the star crust.
  This implies that the core material is just incapable of sustaining Alfv\'en vibrations
  which owe their existence to extremely large (effectively infinite) electrical
  conductivity of matter\cite{Chandra-61,AF-63,M-99}.
  The micro-composition of crust, which is dominated by nuclei embedded 
  in degenerate Fermi- gas of relativistic electrons, suggests that its metal-like material
  possesses properties of perfectly conducting solid-state plasma.
  Such a view suggests that seismic stability of
  the star to quake-induced tectonic displacements of crust against core is primarily
  determined by well-known effect of magnetic (magnet-metal) cohesion mediated
  by magnetic field lines which operate as a super-hard piles
  endowing the core-crust construction of neutron star with supplementary
  (to gravity forces) stiffness of magnetic nature. The intermediate layer between
  core and crust (the inner crust whose density several times less than the core 
  density) is most likely composed of
  quasi-boson matter of paired neutrons. But it is highly unlikely that such quasi-
  boson matter is capable of undergoing phase transition to the 
  Bose-Einstein Condensation (BEC) which is characterized by vanishingly
  small pressure. This suggests that BEC state of paired neutrons, if exist, can only
  insufficiently contribute to the total mass budget of neutron star -- a compact
  object in which pressure of self-gravity is brought to equilibrium by degeneracy
  Fermi-pressure of non-relativistic neutrons in the core and relativistic  electrons
  in the crust. In seismo-dynamics of the paramagnetic neutron star under
  consideration the inner crust is thought of as operating like a lubricant facilitating
  differentially rotational shear displacements of crust relative to much denser matter
  of massive core. From the view point of this core-crust model, the star-quake is
  though of as impulsive release of energy of magnetic core-crust cohesion
  (by means of disruption of magnetic field lines on the core-crust interface)
  resulting in the crust fracturing by revealed magnetic stresses.
  In this paper  we focus, however, not on dynamics of quake, but 
  on the post-quake vibrational relaxation of the star, namely, on 
  node-free torsional oscillations of crustal solid-state plasma about axis of magnetic 
  field frozen in an immobile paramagnetic core.  
  In section 2, a brief outline is given of
  theory of solid-magnetics appropriate for the perfectly conducting viscoelastic
  continuous medium pervaded by a magnetic field.
  In section 3, the spectral formulas for the frequency
  and lifetime of differentially rotational, torsional, nodeless vibrations of the crust
  restored by combined action of magnetic Lorentz and elastic Hooke's forces are obtained.
  In section 4, the computed frequency spectra are used for the forward 
  asteroseismic analysis of the fast oscillations of X-ray outburst from above 
  mentioned magnetars. The obtained results are briefly summarized in section 5.

\section{Governing equations of solid-magnetics}

 It is generally realized today that seismic vibrations of superdense
 matter of non-convective FS-stars (white dwarfs, pulsars and quark stars)
 can be properly described by equations
 of solid-mechanical theory of viscoelastic  continuous   
 media\cite{H-80,M-88,Bl-89,B-99a,B-99b,F-00}. 
 In what follows we deal with the shear differentially rotational fluctuations of viscoelastic
 crustal matter of density $\rho$ which are described by quake-induced material 
  displacements $u_i$ (basic variable of solid-mechanics). The non-compressional   
  character of vibrations under consideration implies 
  that $\delta\rho=-\rho\nabla_k u_k=0$. With this in
 mind, the governing equations of solid-magnetics
 (solid-mechanical counterpart of equations of magneto-fluid-mechanics)
 can be written in the form
  \begin{eqnarray}
  \label{e2.1}
  \rho{\ddot u}_i=\nabla_k\,\tau_{ik}+\nabla_k\,\sigma_{ik}+\nabla_k\,\pi_{ik},\quad \nabla_k u_k=0
 \end{eqnarray}
  presuming that Hooke's elastic stresses $\sigma_{ik}$ and Newton's viscous stresses $\pi_{ik}$ are described
  by linear constitutive equations
  \begin{eqnarray}
 \label{e2.2}
 && \sigma_{ik}=2\mu\,{u}_{ik},\quad {u}_{ik}=\frac{1}{2}[\nabla_i {u}_k+\nabla_k {u}_i],\\
 && \label{e2.3}
 \pi_{ik}=2\eta{\dot u}_{ik},\quad {\dot u}_{ik}=\frac{1}{2}[\nabla_i {\dot u}_k+\nabla_k {\dot u}_i]
 \end{eqnarray}
  where $\mu$ stands for the shear modulus, $\eta$ for shear viscosity and $u_{ik}$
  is the tensor of shear strains or deformations. The central to our further discussion is
  the tensor of fluctuating magnetic field stresses
  \begin{eqnarray}
 \label{e2.4}
 &&\tau_{ik}=\frac{1}{4\pi}[B_i\delta B_k+B_k\delta B_i -B_j\delta B_j \delta_{ik}],\\
 \label{e2.4a}
 &&\delta B_i=\nabla_k\,[u_i\,B_k-u_k\,B_i]
 \end{eqnarray}
 As in our previous works\cite{B-09a,B-97}, we consider model with homogeneous internal
 magnetic field whose components in spherical polar coordinates read
  \begin{eqnarray}
  \label{e2.5}
   && B_r=B\cos\theta,\quad\quad B_\theta=-B\sin\theta,
 \quad\quad B_\phi=0
 \end{eqnarray}
  and external dipolar magnetic field is described by ${\bf B}=\nabla\times {\bf A}$, where
  ${\bf A}=[0, 0, A_\phi={\rm m}_s/r^2]$ is the vector potential with the standard parametrization of the dipole magnetic moment ${\rm m}_s=(1/2)BR^3$ of star of radius $R$ and by $B$ is
  understood the magnetic field intensity at its magnetic poles, $B=B_p$.

\subsection{The energy method}

 This method of computing frequency of shear vibrations rests on the equation of energy balance
 \begin{eqnarray}
  \label{e2.6}
 \frac{\partial }{\partial t} \int \frac{\rho {\dot u}^2}{2}\,
 d{\cal V}=-\int [\tau_{ik}+\sigma_{ik}+\pi_{ik}]\,{\dot u}_{ik}\,d{\cal V}
 \end{eqnarray}
 which is obtained by scalar multiplication of equation of magneto-solid-mechanics (\ref{e2.1}) with ${\dot u}_i$ and integration over the volume of seismogenic layer.
 From the technical point of view,
 the shear character of material distortions brought about by forces under consideration owes its origin to
 the symmetric form of stress-tensors in terms of which these forces are expressed.
 It is this last feature of solid-mechanical elastic stresses and magnetic field stresses  
 that endows the solid-state plasma pervaded by homogeneous magnetic field with 
 the capability of responding to non-compression perturbation  by reversal shear vibrations (which are not 
  accompanied by fluctuations in density). 
  The physical significance and practical usefulness of the energy method
 under consideration is that it can be efficiently utilized not only in the study  
 of non-radial seismic vibrations of neutron stars,  but also can be applied to the study 
 of more wide class of
 solid degenerate stars like white dwarfs stars\cite{L-95, Mol-10} and ultra dense 
 quark-matter stars\cite{W-05} whose material is most likely in the 
 solid aggregate state\cite{Xu-03,Xu-09}. At this point it seems 
 appropriate to mention here theoretical investigations of vibration properties 
 of atomic nuclei (thought of as ultra fine pieces of continuous nuclear matter) 
 in which it has been found that nuclear giant-resonant 
 excitations (fundamental vibration modes generic to all nuclei of periodic chart) are properly described in terms of spheroidal and torsional elastic vibrations of a solid 
 sphere\cite{B-08b}. This suggests that degenerate nucleon Fermi-matter, regarded
 as continuous medium, can be thought of as a strained Fermi-solid, rather than 
 flowing Fermi-liquid.

 The key idea of this method consists in using of the following separable form of material displacements
 \begin{eqnarray}
 \label{e2.7}
 && u_i({\bf r},t) =a_i({\bf r}){\alpha}(t)
 \end{eqnarray}
 where $a_i({\bf r})$ is the time-independent solenoidal field and amplitude ${\alpha}(t)$
 carries information about temporal evolution of fluctuations. Thanks to this form 
 of $u_i$, all the above tensors of fluctuating stresses and strains take similar separable form
 \begin{eqnarray}
  \label{e2.8}
 && \tau_{ik}({\bf r},t)=[{\tilde \tau}_{ik}({\bf r})-\frac{1}{2}{\tilde \tau}_{jj}({\bf r})\delta_{ik}]\alpha(t),\\
  \label{e2.9}
 && {\tilde \tau}_{ik}({\bf r})=\frac{1}{4\pi}
 [B_i({\bf r})\,b_k({\bf r})+B_k({\bf r})\,b_i({\bf r})],\\
 \label{e2.10}
 && b_i({\bf r})=\nabla_k\,[a_i({\bf r})\,B_k({\bf r})-a_k({\bf r})\,B_i({\bf r})],\\
 \label{e2.11}
 && {\sigma}_{ik}({\bf r})=2\mu\,a_{ik}({\bf r})\alpha(t),\quad {\pi}_{ik}({\bf r})=2\eta\,a_{ik}({\bf r}){\dot \alpha}(t),\\
  \label{e2.12}
 && a_{ik}({\bf r})=\frac{1}{2}[\nabla_i a_k({\bf r})+\nabla_k a_i({\bf r})].
 \end{eqnarray}
 On inserting (\ref{e2.7})-(\ref{e2.12}) in the integral equation of energy balance 
 (\ref{e2.6})
 we arrive at equation for $\alpha(t)$ having the well-familiar form
\begin{eqnarray}
 \label{e2.13}
 && \frac{d{\cal E}}{dt}=-2{\cal F},\,\, {\cal E}=\frac{{\cal M}{\dot \alpha^2}}{2}+\frac{{\cal K}\alpha^2}{2},\,\,
 {\cal F}=\frac{{\cal D}{\dot \alpha^2}}{2}\\
 \label{e2.14}
 &&{\cal M}{\ddot \alpha}+{\cal D}{\dot
 \alpha}+{\cal K}\alpha=0,\\
  \label{e2.14a}
 &&\alpha(t)=\alpha_0\exp(- t/\tau)\cos(\Omega t),\\
 \label{e2.15}
 &&\Omega^2=\omega^2\left[1-(\omega\tau)^{-2}\right],\,\,
  \omega^2=\frac{{\cal K}}{{\cal M}},\quad \tau=\frac{2{{\cal M}}}{{\cal D}}.
 \end{eqnarray}
 where  the inertia ${\cal  M}$, viscous friction ${\cal  D}$ and stiffness ${\cal  K}$
 of damped oscillator are given by
 \begin{eqnarray}
  \label{e2.16}
  && {\cal M}=\int \rho({\bf r}) a_i({\bf r})\,a_i({\bf r})\,d{\cal V},\quad {\cal K}={\cal K}_e+{\cal K}_m,\\
 \label{e2.17}
  && {\cal K}_e= 2\int \mu({\bf r})\,a_{ik}({\bf r})\,a_{ik}({\bf r})\,d{\cal V},\\
 \label{e2.18}
 && {\cal K}_m=\int \,{\tilde \tau}_{ik}({\bf r})\,a_{ik}({\bf r})\,d{\cal V},\\
 \label{e2.19}
  && {\cal D}= 2\int \eta({\bf r}) \,a_{ik}({\bf r})\,a_{ik}({\bf r})\,d{\cal V}.
 \end{eqnarray}
 All the above equations valid for arbitrary volume occupied by
 magneto-active solid-state plasma whose density, shear modulus and shear viscosity are arbitrary functions of position. As in our previous works, here we confine our computations to the case of uniform profile of these later parameters. 
And because the purpose of our present 
 study is the frequency spectrum of node-free torsion vibrations about 
 magnetic axis of the star, several comments should be made regarding the 
 axisymmetric field of material displacements ${\bf a}$ which is taken 
 in one and the same shape in computing parameter of inertia ${\cal M}$ and spring constants of both solid-mechanical ${\cal K}_e$ and magneto-mechanical 
 ${\cal K}_m$ stiffness.
 
  From the physical point of view, the main argument 
  justifying the use of one and the same field of material displacements in computing 
  frequency spectra of torsional vibrations driven by forces of elastic and magnetic 
  field stresses rests on general statement of continuum-mechanical theories of 
  magneto-active perfectly conducting continuous media -- magnetic field pervading 
  (both liquid-state and solid-state) plasmas imparts to such a medium 
  a supplementary portion of elasticity which is manifested in its capability 
  of transmitting non-compressional mechanical perturbations by transverse Alfv\'en
  waves. Such a view is substantiated by the commonly known fact that transverse 
  wave in incompressible continuous medium is the 
  feature of material oscillatory behavior which is 
  generic to elastic solid, not an incompressible flowing liquid. 
  The transverse hydromagnetic wave propagating, along the lines  
  of constant magnetic field $B$ frozen-in the perfectly conducting medium, 
  with Alfv\'en speed 
  $v_A=[2P_B/\rho]^{1/2}$ (where $P_B=B^2/8\pi$ is the magnetic field pressure),  
  is characterized by dispersion equation $\omega=v_A\,k$ which is similar to 
  that for transverse wave of shear mechanical displacements,  $\omega=c_t\,k$, 
  traveling in an elastic solid with the speed $c_t=[\mu/\rho]^{1/2}$ where $\mu$ 
  is the shear modulus (which has physical dimension of pressure).
  The profound discussion of analogy between oscillatory behavior 
  of incompressible perfectly conducting plasmas and elastic solid 
  (similarity of transverse hydromagnetic wave in incompressible perfectly conducting 
  media and transverse wave of shear mechanical displacements in 
  an elastic solid) can be found in monographs of Chandrasekhar\cite{Chandra-61} 
  (section Alfv\'en waves), 
  and more extensively this issue is discussed in monograph of Alfv\'en and 
  F\"altammar\cite{AF-63}.  Regarding the difference between node-free 
  oscillatory behavior of 
  solid sphere and spherical mass of an incompressible liquid it is appropriate 
  to note that the liquid sphere 
  is able to sustain solely spheroidal node-free vibrations of fluid velocity. The 
  canonical example is the Kelvin fundamental mode  of oscillating fluid velocity in 
  a heavy spherical mass of incompressible homogeneous  liquid 
  restored by forces represented as gradient of pressure and gradient of  
  potential of self-gravity\cite{B-09c}. 
  In the meantime, the node-free vibrations of solid sphere restored by elastic force 
  (represented as divergence of shear mechanical stresses) are characterized by two 
  eigenmodes. Namely, the even-parity spheroidal mode of non-rotational vibrations of material 
  displacements and the odd-parity torsional mode of differentially rotational vibrations, the problem in which the very 
  notion of the torsion vibration mode has come into existence \cite{B-07a}.
  In our studies focus is laid on poorly investigated regime of node-free  
  vibrations
  in which solenoidal field of material displacements, $\nabla\cdot{\bf a}=0$, obeys 
  the vector Laplace equation, $\nabla^2\,{\bf a}=0$. In this regime 
  the instantaneous material displacements are described by the toroidal field 
  of the form ${\bf a}=A_\ell\nabla\times [{\bf r}r^\ell\,P_\ell(\cos\theta)]$. 
  Substituting this
  field in the above given integrals for inertia ${\cal M}$ and solid-mechanical 
  stiffness ${\cal K}_e$ and integrating over the entire volume of oscillating star we 
  have found in our 
  previous studies that elastic-force-driven node-free global torsional vibrations, ${\cal M}{\ddot\alpha}+{\cal 
  K}_e{\alpha}=0$, are characterized by the frequency spectrum 
  $\omega_e(\ell)=[{\cal K}_e/{\cal M}]^{1/2}$ of the form
    \begin{eqnarray}
 \label{e3.14}
  && \omega_e^2(\ell)=\omega_e^2\left[(2\ell+3)(\ell-1)\right],\\
 && \omega_e=\frac{c_t}{R},\, c_t=\sqrt{\frac{\bar\mu}{\bar\rho}}.
 \end{eqnarray}
 where $\omega_e$ is the natural unit of frequency of shear elastic vibrations and
 $c_t$ is the speed of transverse wave in the elastic solid characterized by 
 average shear modulus  $\bar\mu$ and average density $\bar\rho$. 
  
  The above line of argument about similarity between oscillatory behavior 
  of elastic solid and magneto-active plasma suggests that perfectly conducting 
  matter of neutron star pervaded by homogeneous magnetic field 
  should be able to sustain the Lorentz-force-driven differentially rotational 
  seismic vibrations (triggered by quake) about magnetic axis in which oscillating 
  field of material displacements has one and the same form as in the above outlined  
  Hooke's-force-driven torsional vibrations. Adhering to this assumption  
  and making use of the above node-free toroidal field, 
  ${\bf a}$, as a trial function for computing ${\cal M}$ and ${\cal K}_m$ we 
  found  in\cite{B-09a,B-09b} that torsional vibrations 
  restored by magnetic Lorentz force, ${\cal M}{\ddot\alpha}+{\cal 
  K}_m{\alpha}=0$, are characterized by the frequency spectrum   
  \begin{eqnarray}
 \label{e3.16}
 && \omega_m^2(\ell)=\omega^2_A\left[(\ell^2-1)\frac{2\ell+3}{2\ell-1}\right],\\
 && \omega_A=\frac{v_A}{R},\,\, v_A=\frac{B}{\sqrt{4\pi\bar\rho}}
 \end{eqnarray}   
 where $\omega_A$ is the natural unit of frequency of Alfv\'en vibrations and  $v_A$   
 is the speed of Alfv\'en wave. The practical usefulness of outlined computations with one and the same trial toroidal field of displacements 
 is that allows us to assess the relative role of restoring Hooke's and Lorentz forces 
 in torsional seismic vibrations of neutrons stars which 
 are responsible, as is believed, for the fast oscillations of X-ray flares  
 from quaking magnetars.  With all above in mind, one of the 
  main purposes of present paper is to make such an assessment 
  in a mathematically consistent fashion for the torsional node-free 
  vibrations entrapped in the crust, the problem which is considered,  to the best of 
  our knowledge, for the first time.

\subsection{Material displacements in torsional mode of crust-against-core
nodeless vibrations}

  The above equations of the energy method show that the 
  main trial function of the frequency spectrum computation 
  is the toroidal field of instantaneous displacements. Because one of the main 
  our purposes here is to assess the relative role of elastic and magnetic forces in 
  quake-induced torsion nodeless vibrations of magnetars, 
  we again adopt all the above arguments regarding the choice of this field in the 
  form of the general solution to vector Laplace equation. 
  
  It is convenient to start with the rate of 
  material displacements which is described by general formula of rotational motions
  \begin{eqnarray}
  \label{e3.1}
  && \delta {\bf v}({\bf r},t)={\dot {\bf u}}({\bf r},t)=[\mbox{\boldmath $\Omega$}({\bf r},t)\times {\bf r}],\\
  \label{e3.2}
  && \mbox{\boldmath $\Omega$}({\bf r},t)=
  [\nabla\times \delta {\bf v}({\bf r},t)]=[\nabla\times {\dot {\bf u}}({\bf r},t)]={\dot{ \mbox{\boldmath $\Phi$}}}({\bf r},t)
  \end{eqnarray}
   However, unlike a case of rigid-body rotation, in which the angular velocity is a constant vector,
   in a solid mass undergoing axisymmetric differentially rotational vibrations the angular velocity $\mbox{\boldmath $\Omega$}({\bf r},t)$ is the vector-function of position which
   can be represented as
 \begin{eqnarray}
  \label{e3.3}
  && \mbox{\boldmath $\Omega$}({\bf r},t)={\dot{ \mbox{\boldmath $\Phi$}}}({\bf r},t)=\mbox{\boldmath $\phi$}({\bf r})\,{\dot \alpha}(t),\,\, \mbox{\boldmath $\phi$}({\bf r})=[\nabla\times {\bf a}({\bf r})]
  \end{eqnarray}
  In the regime of node-free vibrations in question, ${\bf a}({\bf r})$ is described by the divergence-free odd-parity, axial, toroidal field which is one of two harmonic solenoidal fields of fundamental basis\cite{Chandra-61}
  obeying the vector Laplace equation $\nabla^2 {\bf a}=0$. This field can be expressed
  in terms of general solution of the scalar Laplace equation as follows
   \begin{eqnarray}
 \label{e3.4}
 && {\bf a}({\bf r})={\bf a}_t({\bf r})=\nabla \times [{\bf r}\,\chi({\bf r})]=
 [\nabla \chi({\bf r}) \times {\bf r}]\\
  \label{e3.5}
 && \nabla^2\chi({\bf r})=0,
 \\
 && \chi({\bf r})=[A_\ell\,r^\ell+B_\ell\,r^{-\ell-1}]\,P_\ell(\zeta),\quad \zeta=\cos\theta
  \end{eqnarray}
 where $P_\ell(\zeta)$ is the Legendre polynomial of multipole degree $\ell$. It follows that the angular field $\mbox{\boldmath $\phi$}({\bf r})$ is the poloidal vector field
  \begin{eqnarray}
 \label{e3.6}
 \mbox{\boldmath $\phi$}({\bf r})&=&[\nabla\times {\bf a}_t({\bf r})]=\nabla\times \nabla\times [{\bf r}\,\chi({\bf r})]\\
 &=&\nabla\,[{\cal A}_\ell\,r^\ell+{\cal B}_\ell\,r^{-\ell-1}]\,P_\ell(\zeta),\\
 && {\cal A}_\ell=A_\ell(\ell+1),\quad {\cal B}_\ell=B_\ell\,\ell.
 \end{eqnarray}
 And this field is irrotational, $\nabla\times \mbox{\boldmath $\phi$}({\bf r})=0$.

\begin{figure}[ht]
\centering{\includegraphics[width=7.0cm]{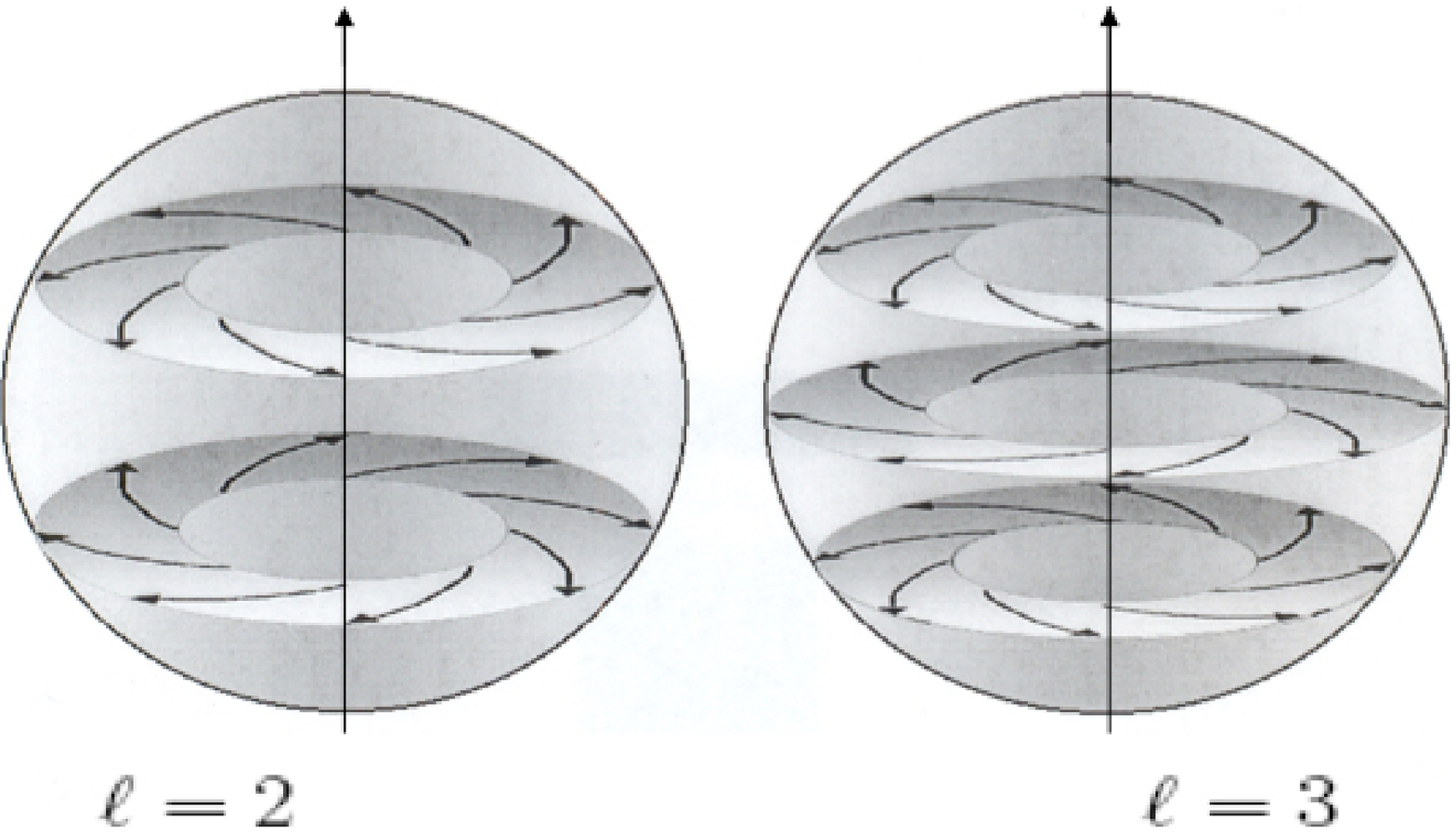}}
 \caption{(Color online) Material displacements of crustal matter about the dipole magnetic moment axis of paramagnetic neutron star undergoing nodeless differentially rotational, torsional, vibrations in quadrupole and octupole overtones.}
\end{figure}

 As was stated, we study a model of differentially rotational vibrations of peripheral finite-depth crust against
 immobile core.  In this case, the arbitrary constants ${\cal A}_\ell$ and ${\cal B}_\ell$ can be
 uniquely eliminated from two boundary conditions: (i) on the core-crust interface $r=R_c$
 \begin{eqnarray}
   \label{e3.7}
 u_\phi\vert_{r=R_c}=0
 \end{eqnarray}
and (ii) on the star surface $r=R$
  \begin{eqnarray}
   \label{e3.8}
 && u_{\phi}\vert_{r=R}=[\mbox{\boldmath $\Phi$}\times {\bf R}]_\phi\vert_{r=R},
 \\
 && \mbox{\boldmath $\Phi$}=\alpha(t)\nabla_{\hat n} P_\ell(\zeta),\quad {\bf R}={\bf e}_rR
 \end{eqnarray}
 where
 \begin{eqnarray}
 \label{e3.9}
  \nabla_{\hat n}=\frac{1}{R}\nabla_{\mbox{\boldmath $\Omega$}},\quad \nabla_{\mbox{\boldmath $\Omega$}}= \left[{\bf e}_\theta \frac{\partial }{\partial
 \theta}+{\bf e}_\phi\frac{1}{\sin\theta}\frac{\partial }{\partial \phi}\right].
 \end{eqnarray}
 The no-slip condition on the core-crust interface, $r=R_c$,
 reflects the fact that the amplitude of differentially rotational oscillations
 gradually decreases down to the star center and turns into zero on the core.
 The boundary condition on the  star surface, $r=R$,
 is dictated by symmetry of the general toroidal solution of the vector Laplace equation which then
 is tested to reproduce the moment of inertia of a rigidly rotating solid star\cite{B-99b, B-08}. The above boundary conditions lead to the coupled algebraic equations
 \begin{eqnarray}
 \label{e3.10}
&&{\cal A}_\ell R_c^{\ell-1}+{\cal B}_\ell R_c^{-\ell-2}=0,\,\, {\cal A}_\ell R^{\ell}+{\cal B}_\ell R^{-\ell-1}=R
\end{eqnarray}
whose solutions are
\begin{eqnarray}
  \label{e3.11}
 {\cal A}_\ell={\cal N}_\ell,\,{\cal B}_{\ell}=-{\cal
 N}_\ell\,R_c^{2\ell+1},\,\,{\cal
 N}_\ell=\frac{R^{\ell+2}}{R^{2\ell+1}-R_c^{2\ell+1}}.
 \end{eqnarray}
In spherical polar coordinates, the nodeless toroidal field has only one non-zero azimuthal component
\begin{eqnarray}
 \label{e3.12}
 && a_{r}=0,\,\, a_{\theta}=0,\\
 \nonumber
 && a_{\phi}=\left[{\cal A}_\ell\,r^\ell+\frac{{\cal B}_\ell}{r^{\ell+1}}\right](1-\zeta^2)^{1/2}\frac{d P_\ell(\zeta)}{d\zeta}.
 \end{eqnarray}
 The snapshot of material node-free displacements in the crust undergoing torsional oscillations against immobile core of paramagnetic neutron star under consideration is pictured in
 Fig.2 for
 quadrupole, $\ell=2$, and octupole $\ell=3$, overtones of this axial mode.
 The adopted first boundary 
   condition (\ref{e3.7}) implying that all stresses  (elastic, magnetic and viscous) 
   vanish on the core-crust interface suggests that quake-induce perturbation sets in 
   the node-free torsional motions
  only a finite-depth crustal region, whereas central undisturbed 
  region of the star remains at rest.  In the next section 
  the result of analytic computations are presented in the form showing 
  that spectral formulas for toroidal modes entrapped in the crust are reduced 
  to the above presented ones, equations (\ref{e3.14}) and (\ref{e3.16}), for the global oscillations (in the entire volume of the 
  star) when core radius 
  tends to zero; this fact is regarded as a test justifying mathematical correctness of 
  presented computations.  
 
\section{Spectral formulae for the frequency and lifetime}

 The computation of integrals defining mass parameter ${\cal M}$, parameter of vibrational rigidity ${\cal K}$ and viscous friction ${\cal D}$ [which has been presented in some details elsewhere\cite{B-99b,B-07a} are quite lengthy but straightforward and, therefore,
 are not presented here. The mass parameter can be conveniently
 represented in the form
\begin{eqnarray}
 \label{e4.1}
 && {\cal M}=4\pi\,\rho R^5\frac{\ell(\ell+1)}{(2\ell+1)(2\ell+3)}\,m(\ell),\\
 \label{e4.1a}
 && m(\ell)=(1-\lambda^{2\ell+1})^{-2}\\
 \nonumber
 && \times \left[1-
 (2\ell+3)\lambda^{2\ell+1}
  \right. +\\ \nonumber
 && +  \left.
 \frac{(2\ell+1)^2}{2\ell-1}\lambda^{2\ell+3}-
 \frac{2\ell+3}{2\ell-1}\lambda^{2(2\ell+1)}\right],\\
 \label{e4.2}
 && \lambda=\frac{R_c}{R}=1-h,\quad h=\frac{\Delta R}{R},\\
 \label{e4.2a}
 && \Delta R=R-R_c,\quad 0\leq \lambda <1.
\end{eqnarray}
The $\lambda$-terms in the above and foregoing equations emerge as result of integration
along the radial coordinate from radius of the core-crust interface $r=R_c$ to the star radius, $r=R$.
The integral coefficient of viscous friction is given by
\begin{eqnarray}
\label{e4.3}
 && {\cal D}=4\pi\eta R^3\frac{\ell(\ell^2-1)}{2\ell+1}
 \,d(\ell),\\
 \label{e4.4}
 && d(\ell)={(1-\lambda^{2\ell+1})^{-1}}
 \left[1-\frac{(\ell+2)}{(\ell-1)}\lambda^{2\ell+1}\right].
\end{eqnarray}

 \begin{figure}[ht]
\centering{\includegraphics[width=8.cm]{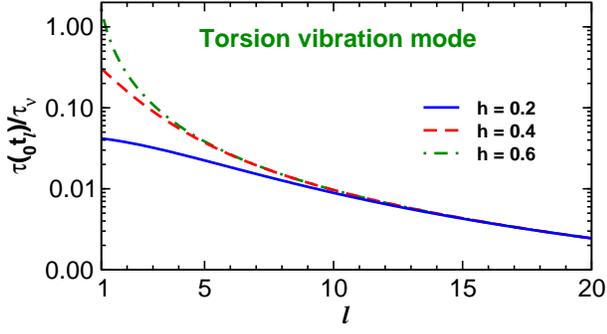}}
 \caption{(Color online) The fractional lifetime of torsion nodeless oscillations of the neutron star crust damped by force of viscous shear stresses as a function of multipole degree $\ell$ computed at indicated values of the fractional depth $h$ of peripheral seismogenic layer.}
\end{figure}

For the lifetime we obtain
 \begin{eqnarray}
 \label{e4.5}
 && \tau(_0t_\ell)=\frac{2\tau_\nu}{(2\ell+3)(\ell-1)}\,\frac{m(\ell)}{d(\ell)},
 \,\, \tau_\nu=\frac{R^2}{\nu},\,\,\nu=\frac{\eta}{\rho}.
\end{eqnarray}
 In Fig.3, the fractional lifetime
 is plotted as a function of multipole degree $\ell$ with indicated values of fractional depths $h=\Delta R/R$.
 It shows that the higher $\ell$ the shorter lifetime. It is easy to see that in the limit, $\lambda=(R_c/R)\to 0$, we regain the spectral formula for lifetime of global torsional nodeless vibrations of solid star\cite{B-03,B-07a}
\begin{eqnarray}
  \label{e4.6}
 && \tau(_0t_\ell)=\frac{2{\bar\tau}_\nu}{(2\ell+3)(\ell-1)},\quad {\bar \tau}_\nu=\frac{R^2}{\bar\nu},\quad {\bar \nu}=\frac{\bar\eta}{\bar\rho}.
 \end{eqnarray}
 in which by $\bar\tau_\nu$ is understood, in this latter case,  the average kinematic viscosity
 of the star matter as a whole; the extensive discussion of this transport coefficient can be found in\cite{S-08}. For the node-free torsional oscillations of solid star, the last equation
 has one and the same physical significance as the well-known Lamb formula does for the
 time of viscous damping of spheroidal node-free vibrations which
 in the context of neutron star pulsations has been extensively discussed in\cite{CL-87}.
 Regarding the problem under consideration we cannot see, however, how
 the obtained formulae can be applied to observational data on QPOs in SGRs.
 Nonetheless, their practical usefulness is that they can be utilized in the study
 of a more wide class of solid celestial objects such as Earth-like planets\cite{LW-95,AR-02} and white dwarf stars\cite{HKT-04}.

 From above it is clear that the integral coefficient of elastic rigidity ${\cal K}_e$ of torsional vibrations
 has analytic form similar to that for coefficient of viscous friction ${\cal D}$, namely
\begin{eqnarray}
\label{e4.7}
 && {\cal K}_e=4\pi\mu R^3\frac{\ell(\ell^2-1)}{2\ell+1}
 \,k_e(\ell),\\
\label{e4.8}
 && k_e(\ell)={(1-\lambda^{2\ell+1})^{-1}}
 \left[1-\frac{(\ell+2)}{(\ell-1)}\lambda^{2\ell+1}\right].
\end{eqnarray}
The frequency as a function of multipole degree  $\ell$  of node-free elastic vibrations in question $\nu_e(\ell)$ (measured in Hz and related to angular frequency as $\omega_e(\ell)=2\pi\nu_e(\ell)={\cal K}_e/{\cal M}$) is given by
 \begin{eqnarray}
 \label{e4.9}
 && \nu_e^2(\ell)=\nu^2_e\left[(2\ell+3)(\ell-1)\right]\,
 \frac{k_e(\ell)}{m(\ell)},\\
 \label{e4.10}
 && \omega_e=2\pi\nu_e=\frac{c_t}{R},\, c_t=\sqrt{\frac{\mu}{\rho}},\,
 \lambda=1-h,\, h=\frac{\Delta R}{R}.
 \end{eqnarray}
 It is easy to see that in the limit $\lambda\to 0$,  we regain spectral formula for the frequency of global torsional oscillations, having the form of equation (\ref{e4.9}) with $k_e(\ell)=m(\ell)=1$. Understandably that in this latter case all material characteristics belong to the star as a whole.

\begin{figure}[ht]
\centering{\includegraphics[width=7.5cm]{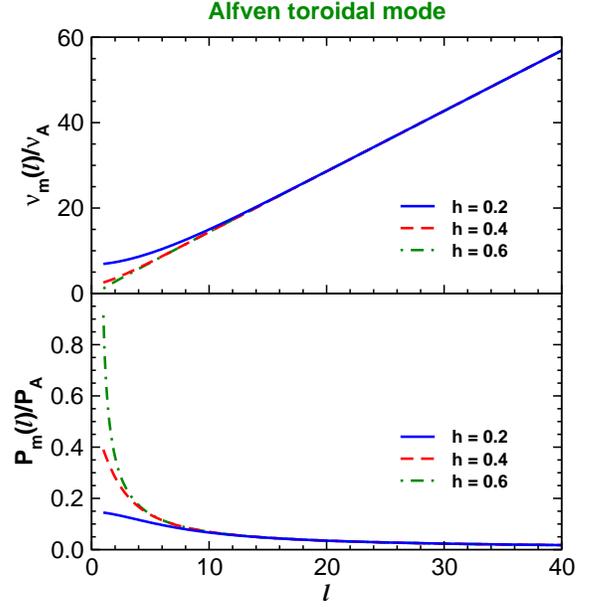}}
 \caption{(Color online) Fractional frequency and period of nodeless torsional magneto-solid-mechanical oscillations,
 toroidal Alfv\'en mode -- $_0a^t_\ell$, entrapped in the neutron star crust as functions of multipole degree $\ell$
 computed at indicated values of the fractional depth $h$ of peripheral seismogenic layer. The value
 $h=1$ corresponds to global torsional oscillations excited in the entire volume of the star. Here $\nu_A=\omega_A/2\pi$, where $\omega_A=v_A/R$ with $v_A=B/\sqrt{4\pi\rho}$ being the velocity
 of Alfv\'en wave in crustal matter of density $\rho$ and $P_A=2\pi/\omega_A$.}
\end{figure}

 The magneto-mechanical stiffness of Alfv\'en vibrations ${\cal K}_m$
 can conveniently be written as
\begin{eqnarray}
 \label{e4.11}
 && K_m=B^2R^3\frac{\ell(\ell^2-1)(\ell+1)}{(2\ell+1)(2\ell-1)}\,k_m(\ell)\\
 \label{e4.12}
 && k_m(\ell)=(1-\lambda^{2\ell+1})^{-2}\, \left\{1+\frac{3\lambda^{2\ell+1}}{(\ell^2-1)(2\ell+3)}
  \right. \times \\ \nonumber
 && \times \left.
 \left[1-\frac{1}{3}\ell(\ell+2)(2\ell-1)
 \lambda^{2\ell+1}\right]\right\}.
 \end{eqnarray}
 This leads to the following two-parametric spectral formula
 \begin{eqnarray}
 \label{e4.13}
 && \nu_m^2(\ell)=\nu^2_A\left[(\ell^2-1)\frac{2\ell+3}{2\ell-1}\right]\,
 \frac{k_m(\ell)}{m(\ell)},\\
 \label{e4.14}
 && \omega_A=2\pi\nu_A=\frac{v_A}{R},\quad v_A=\frac{B}{\sqrt{4\pi\rho}}.
 \end{eqnarray}
 In the forward asteroseismic analysis of QPO data
 relying on this latter spectral formula, the  Alfv\'en frequency, $\nu_A$ and the fractional depth of seismogenic zone, $h$, are regarded as free parameters which are adjusted so as to reproduce general trends in the observed
 QPOs frequencies. 
  In Fig.4, the fractional frequencies and periods of this toroidal Alfv\'en mode
  as functions of multipole degree $\ell$
 are plotted with indicated values of fractional depth of the seismogenic layer $h$.
 Remarkably, the lowest overtone of global oscillations is of quadrupole degree, $\ell=2$, whereas for
 vibrations locked in the crust, the lowest overtone is of dipole degree, $\ell=1$, as is clearly seen
 in Fig.5. This suggests that dipole vibration can be thought of as
 Goldstone's soft mode whose most conspicuous property is that the mode
 disappears (the frequency tends to zero) when key parameter regulating the depth of seismogenic zone $\lambda\to 0$.

\begin{figure}[ht]
\centering{\includegraphics[width=7.5cm]{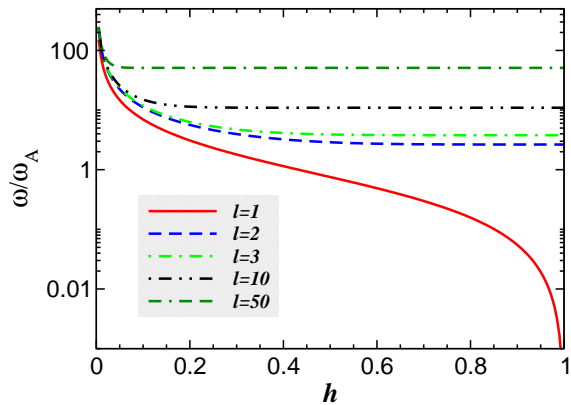}}
 \caption{(Color online) Fractional frequency of nodeless torsional Alfv\'en oscillations
 of indicated overtones $\ell$ as a function of the fractional depth $h$ of peripheral seismogenic layer.
 The vanishing of dipole overtone in the limit of $h\to 1$, the case when entire mass of neutron star sets in torsional oscillations, suggests that dipole vibration possesses property typical to Goldstone's soft modes .}
\end{figure}

\section{Application to QPOs in the outburst X-ray flux from SGR 1806-20 and SGR 1900+14}

  The basic physics underlying current understanding of interconnection between quasi-periodic oscillations of detected electromagnetic flux and vibrations of neutron star has been recognized long ago\cite{H-80,Bl-89}. Owing to the effect of strong flow-field coupling, which is central to
  the propagation of Alfv\'en waves, the quake induced perturbation excites coupled vibrations of perfectly conducting solid-state plasma of the crust (as well as gaseous plasma of magnetar corona expelled from the surface by outburst) and frozen-in lines of magnetic field. Outside the star the vibrations
  of magnetic field lines are coupled with oscillations of gas-dust plasma expelled from the star surface by
  quake. And it is these fluctuations of outer lines of magnetic field, operating like transmitters of beams of charged particles producing coherent (curvature and/or synchrotron) high-energy radiation, are detected as QPOs
  of  light curves of the SGRs giant flares.

  In applying the obtained spectral formulae to the frequencies of detected QPOs we examine two scenarios, namely, when quake-induced torsional vibrations are restored by joint action of Lorentz magnetic and Hooke's
  elastic forces and when oscillations are of pure Alfv\'en's nature, that is, produced by torsional seismic
  vibrations of crust against core under the action of solely one Lorentz force of magnetic field stresses.

\subsection{Crust vibrations driven by combined action of Lorentz magnetic and Hooke's elastic forces}

In this case, the asteroseismic analysis of detected QPOs rests on the three-parametric spectral formula
   \begin{eqnarray}
 \label{e5.1}
 && \nu^2(\ell)[\nu_A,\nu_e,h]=\nu_m^2(\ell)[\nu_A,h]+\nu^2_e(\ell)[\nu_e,h].
 \end{eqnarray}
  The suggested theoretical $\ell$-pole specification of the detected frequencies
  is presented in Fig. 6 for SGR 1900+14 and in Fig. 7 and Fig. 8, exhibiting remarkable correlation between depth  of seismogenic zone and fundamental frequencies of magnetic and elastic oscillations - the larger $\Delta R$, the higher basic frequencies of Alfv\'enic $\nu_A$ and elastic $\nu_e$ vibrations. It is seen from computations
  for SGR 1806-20, that  reasonable fit of data can be attained with $h=0.2$ (for the star model with radius 20 km, $\Delta R=2$ km) and with $h=0.4$ ($\Delta R=5$ km).

   \begin{figure}[ht]
 \centering{\includegraphics[width=7.5cm]{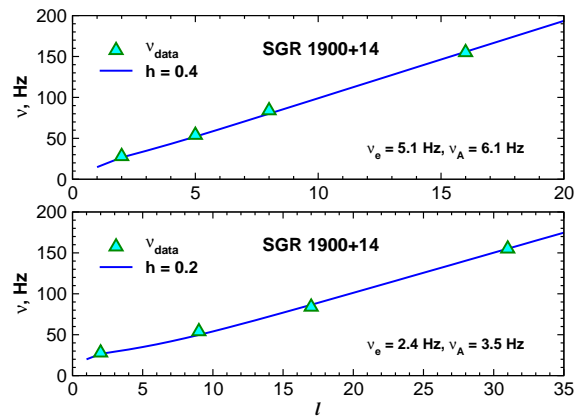}}
 \caption{(Color online) Theoretical fit of the QPOs frequency in the X-ray flux from SGR 1900+14 on the basis
 of three-parametric theoretical spectrum of frequency of torsional seismic
 vibrations in the crustal region of indicated fractional depth.}
\end{figure}

    \begin{figure}[ht]
 \centering{\includegraphics[width=8.0cm]{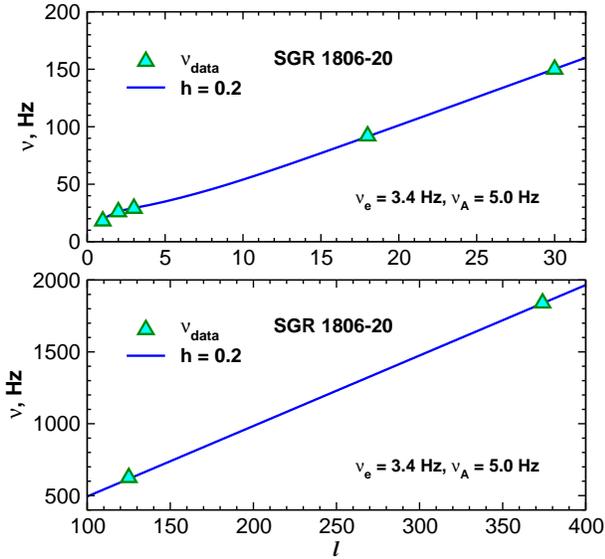}}
 \caption{(Color online) The same as Fig.6, but for SGRs 1806-20 with h=0.2.}
\end{figure}

 \begin{figure}[ht]
 \centering{\includegraphics[width=8.0cm]{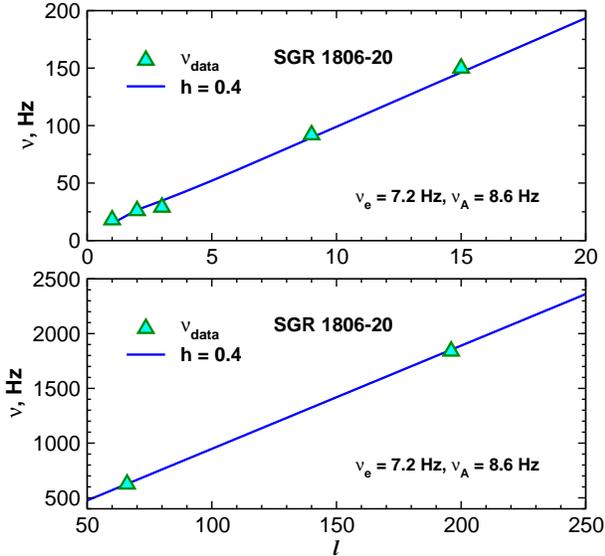}}
 \caption{(Color online) The same as Fig.6, but for  SGRs 1806-20 with h=0.4.}
\end{figure}

  It is worth emphasizing that at above values of $h$, the obtained here tree-parametric spectral formula much better match the data as compared to that for global, in the entire volume, vibrations studied in\cite{B-09a}.
  On this ground we conclude, if the detected QPOs are produced by seismic vibrations of peripheral region of the star under coherent action of Lorentz and Hooke's forces, then the depth of seismogenic layer $\Delta R$ should be quite large, somewhere in the range $0.2R < \Delta R < 0.4R$.

\subsection{Lorentz-force-driven vibrations of crustal solid-state plasma}

  It seems appropriate to note, that pure Alfv\'en oscillations of crustal electron-
  nuclear solid-state plasma about axis of magnetic field frozen in the immobile core
  have been studied some time ago\cite{B-97} in the context
  of searching for fingerprints of post-glitch vibrational behavior of radio pulsars.
  In the problem under consideration, one can use one and the same spectral
  formula for the $\ell$-pole specification of detected QPOs which in above notations
  is written as
 \begin{eqnarray}
 \label{e5.2}
 && \nu^2(\ell)=\nu_m^2(\ell)[v_A,h].
 \end{eqnarray}

  \begin{figure}[ht]
 \centering{\includegraphics[width=8.0cm]{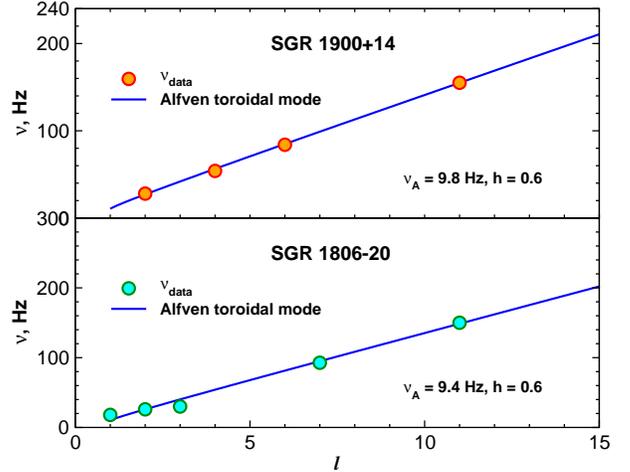}}
 \caption{(Color online) Theoretical description (lines) of detected QPO frequencies (symbols) in the X-ray flux during the flare of SGRs 1806-20 and SGR 1900+14 as overtones of pure Alfv\'en torsional nodeless oscillations of crustal magneto-active plasma under the action of solely Lorentz restoring force.}
\end{figure}

   The results presented in Fig.9 and Fig.10 show that at indicated input parameters, i.e.,
   the Alfv\'en frequency $\nu_A$ and the fractional depth of seismogenic layer $h=\Delta R/R$,
   the model too adequately reproduces general trends in the data with fairly
   reasonable $\ell$-pole specification of overtones pointed out by integer numbers along x-axis.
   It is seen that the low-frequency QPOs in data for SGR 1806-20, are interpreted as dipole and quadrupole overtones: $\nu(_0a^t_1)=18$ and $\ell(_0a^t_2)=26$ Hz. And the high-frequency kilohertz vibrations with $627$ Hz and $1870$ Hz are unambiguously specified as high-multipole overtones: $\nu(_0a^t_{\ell=42})=627$ Hz and $\nu(_0a^t_{\ell=122})=1870$ Hz. However, in this latter scenario of Lorentz-force-dominated vibrations the best fit of data is attained at fairly large value of fractional depth, $h=0.6$, which is much larger than the expected depth of the crust. In our opinion, this result may be regarded as indication to that the detected QPOs are formed by coherent vibrations of crustal solid-state plasma and plasma of magnetar corona.

   \begin{figure}[ht]
 \centering{\includegraphics[width=9.0cm]{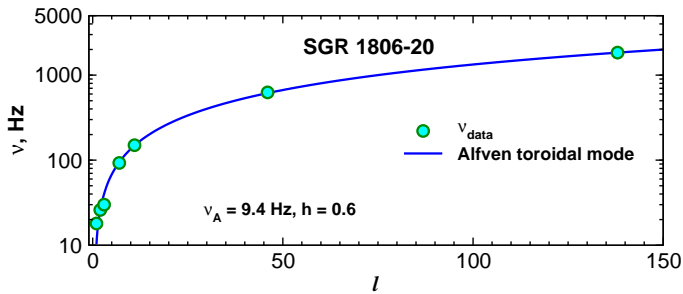}}
 \caption{(Color online) Same as Fig.9 but for SGRs 1806-20.}
\end{figure}

\section{Concluding remarks}

 Ever since identification of pulsars with rapidly rotating neutron stars
 it has been argued\cite{A-81, S-91}  that two key properties of
 these compact objects  -- (i) the degeneracy of neutron
 (non-conducting) Fermi-matter whose pressure opposes
 the pressure of self-gravity, and (ii)
 a highly stable to decay super-strong magnetic fields --
 can be reconciled, if poorly conducting neutron-dominated stellar matter, 
 constituting the neutron star cores, has been brought to gravitational equilibrium in 
 the permanently magnetized state. The most plausible is the state of Pauli's 
 paramagnetic saturation with spin 
 magnetic moments of neutrons polarized along the axis of fossil field 
 inherited from massive progenitor and amplified in magnetic-flux-conserving core-
 collapse supernova\cite{B-02b}. This idea is central to 
 the considered two-component, core-crust, model of paramagnetic neutron star 
 whose less dense and highly conducting, metal-like, material of the crust is 
 considered as a solid-state, electron-nuclear, plasma pervaded by frozen in the core 
 magnetic field.  This difference between electrodynamic properties of core and crust 
 matter (permanently magnetized non-conducting core and perfectly conducting non-
 magnetic crust) suggests that magnetic cohesion between massive core (permanent magnet)
and crust (metal-like material) should plays central role in seismic activity 
of the star. Working from such an understanding, we have computed frequency 
 spectra of node-free torsional oscillations of crust 
 against immobile core under the action  Lorentz and Hooke restoring 
 forces and damped by Newtonian viscous force. As a trial function of oscillating  
 material displacements we have used the node-free toroidal field computed from 
 vector Laplace equation. The obtained spectral formulas are of some 
 interest in their own right because they can be applied to more wide class of 
 celestial objects.  
 In this work we applied the obtained analytic frequency spectra to magnetars, highly 
 magnetized quaking neutron stars whose bursting seismic activity is commonly 
 associated with release of magnetic field stresses. Focus was laid on 
 forward asteroseismic analysis of fast X-ray flux oscillations during the giant flare of 
 SGR 1900+14 and SGR 1806-20 and, thus, assuming that 
 these oscillations are produced by torsion vibrations of crustal solid-state plasma  
 about axis of dipole magnetic field 
 frozen in the immobile permanently magnetized core. 
 In so doing we have investigated two cases of post-quake vibrational relaxation of  
 the star, depending on restoring forces. 
 In first case, the analysis of data has been based on
 assumption that detected QPOs owe their existence to node-free torsional 
 vibrations of crust against core restored by joint action of Lorentz magnetic and 
 Hooke's elastic forces. And we found that obtained three-parametric spectral 
 formula provides much better fit of data than two-parametric frequency spectrum of 
 global vibrations (Bastrukov et al 2009a). The considered second scenario 
 presumes that vibrations are dominated by solely Lorentz restoring force of 
 magnetic field stresses. We found that obtained two-parametric frequency spectrum 
 can too be fairly reasonably reconciled with detected QPOs frequencies.
 All the above lead us to conclude that  Lorentz restoring force of magnetic field 
 stresses plays decisive part in quake-induced torsional vibrations of crustal solid-state plasma of magnetars.
 
  The authors are grateful to Dima Podgainy (JINR, Dubna)  for helpful assistance 
 and referee and reviewer for suggestions clarifying the subject and result of presented investigation.

\end{document}